\documentclass[twocolumn,aps,prl,epsf]{revtex4}
\usepackage{graphicx}
\begin{document}

\draft

\title{Aharonov-Bohm Oscillations with Spin: Evidence for Berry's Phase}

\author{Jeng-Bang Yau, E. P. De Poortere, M. Shayegan}
\address{Department of Electrical Engineering, Princeton University, Princeton, New Jersey 08544}
\date{\today}
\begin{abstract}
We report a study of the Aharonov-Bohm effect, the oscillations of the resistance of a mesoscopic ring as a function of a perpendicular magnetic field, in a GaAs two-dimensional hole system with a strong spin-orbit interaction. The Fourier spectra of the oscillations reveal extra structure near the main peak whose frequency corresponds to the magnetic flux enclosed by the ring. A comparison of the experimental data with results of simulations demonstrates that the origin of the extra structure is the geometric (Berry) phase acquired by the carrier spin as it travels around the ring.
\end{abstract}
\pacs{73.63.-b, 03.65.Vf, 71.70.Ej}
\maketitle

An important and, at times, mysterious concept in modern physics is the phase factor that a quantum mechanical wave function acquires upon a cyclic evolution. This phase factor can lead to interference phenomena which are experimentally observable. An example is the Aharonov-Bohm (AB) effect \cite{AharonovBohm59,Review}, the oscillations in the resistance of a mesoscopic conducting ring as a function of an external magnetic flux piercing the ring. The origin of the oscillations is the phase acquired by the electron wave as it travels around the ring, and the interference of this wave with itself. The phase in this case is equal to $2\pi (\Phi_{ext}/\Phi_{0})$, where $\Phi_{ext}=\pi r^{2} \cdot B_{ext}$ and $\Phi_{0}=(h/e)$ is the flux quantum ($r$ is the ring radius and $B_{ext}$ is the external perpendicular magnetic field). As a result, the resistance of the ring exhibits oscillations periodic in $B_{ext}$ with a frequency equal to $\pi r^{2}/(h/e)$. A requirement for the observation of such oscillations is of course that the electron motion around the ring be phase coherent.

In a seminal paper \cite{Berry84}, M. V. Berry showed that, even in the absence of electromagnetic fields, a quantum state undergoing an adiabatic evolution along a closed curve in parameter space develops a phase which depends only on this curve \cite{AnyCyclic}. Thanks to its fundamental origin, this so-called geometric (or Berry) phase has attracted considerable attention \cite{Shapere89}. However, its experimental observation has been scarce. Evidence for Berry's phase was obtained early on in experiments with neutrons, fiber optics, and quadrupole resonance of nuclei \cite{Shapere89}, but its observation in a condensed-matter system has proved challenging. 

To observe Berry's phase in an electronic system with spin, Loss {\it et al.\/} \cite{Loss90} proposed to study transport in a mesoscopic ring structure in the presence of an orientationally inhomogeneous (e.g., radial) magnetic field. This can be experimentally implemented via fabricating the ring from a material with spin-orbit (SO) interaction. In recent, pioneering studies \cite{Morpurgo98,Nitta99b}, the AB oscillations were studied in an InAs two-dimensional (2D) electron system with strong SO interaction. The Fourier transforms of over 30 traces of AB oscillations were averaged and a small splitting of the main peak in the final Fourier spectrum was interpreted as a possible manifestation of the spin Berry phase. Here we report AB measurements on a GaAs 2D hole system with well-characterized SO interaction \cite{Lu98b}. The Fourier spectra of the AB oscillations contain extra structure, often in the form of side peaks, near the central peak which occurs at $\pi r^{2}/(h/e)$. The shape of this extra structure evolves with the range of magnetic field over which the spectra are taken. We compare this evolution with the results of a realistic simulation which includes Berry's phase in AB oscillations of a system with SO interaction. The comparison provides a striking demonstration of Berry's phase.

The starting material for our experiment is a modulation-doped GaAs/AlGaAs heterostructure, with a 2D hole system (density $\sim$ 2.4 $\times$ 10$^{15}$ m$^{-2}$ and mobility $\sim$ 30 m$^{2}$/ V s) at a distance of 100 nm below the surface. Using standard optical and electron-beam lithography techniques and wet etching, we fabricated the ring structure whose micrograph is shown in Fig. 1(a). The inner and outer radii of the ring have nominal values of 0.475 and 0.725 $\mu$m, respectively. The mean-free-path is about 2 to 3 $\mu$m for the 2D holes in an unpatterned region of the sample. However, because of the narrow width of the ring's arms and surface depletion, a front gate is needed to populate the arms with carriers. It is therefore not possible for us to precisely know the mean-free-path of the holes in the ring. We measured the resistance of the ring at a temperature of about 20 mK in a dilution refrigerator and as a function of magnetic field perpendicular to the plane.

Figure 1(b) shows an example of the measured magnetoresistance of the ring. The resistance, after subtraction of a smooth background [see Fig. 1(c)], reveals clear AB oscillations with an amplitude of $\sim$ 5 $\Omega$ \cite{UCF}. Our key result is that these oscillations are not at a single frequency: as shown in Fig. 2(a), the Fourier transform (FT) of the oscillations exhibits extra structure whose form depends on the range of the magnetic field over which the signal is analyzed. In the remainder of the paper, we will demonstrate that the extra structure is a manifestation of Berry's phase in a system with SO interaction.

\begin{figure}
\centering
\includegraphics[scale=0.325]{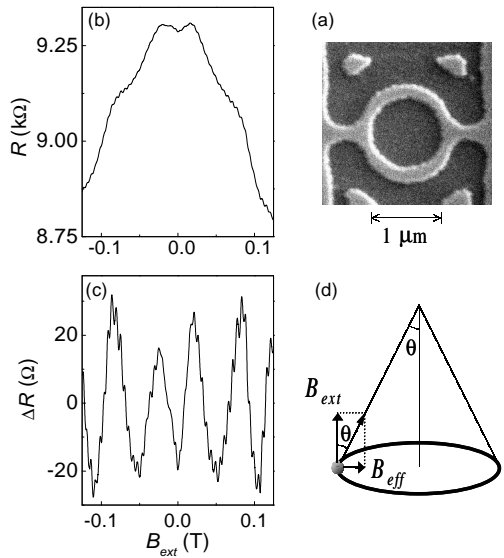}
\vspace{0.1cm}
\caption{(a) Scanning electron microscope picture of the ring structure. (b) A typical trace of the Aharonov-Bohm oscillations measured in GaAs 2D holes at front gate voltage of $-0.7$ V. (c) Data in (b) after subtraction of a smooth background signal. (d) Schematic diagram of a carrier, in the presence of an external magnetic field, travelling around a ring structure in a system with spin-orbit interaction. Its spin precesses around the direction of the total magnetic field {(\bf $\vec{B}_{ext}$+$\vec{B}_{eff}$\/)}.}
\label{Fig1}
\end{figure}

An important and relevant characteristic of the GaAs 2D hole systems is a strong SO interaction which, combined with the inversion asymmetry of the confinement potential, leads to significant spin-splitting of the energy bands in the absence of an applied magnetic field \cite{Lu98b,Bychkov84}. The inversion asymmetry stems partly \cite{BulkInv} from an electric field, which is perpendicular to the 2D plane. In its rest frame, a moving carrier in such systems feels an effective in-plane magnetic field ($B_{eff}$) which is determined by the vector product of the carrier's velocity and this electric field. The field $B_{eff}$ couples to the carrier's spin so that the energy bands at any nonzero wave vector are split into two spin subbands. As a result, the Fermi wave vectors of the opposite-spin carriers occupying the two spin subbands differ by a finite value, $\Delta k$. As we show below, $B_{eff}$ and $\Delta k$ are the key parameters that allow us to demonstrate the observation of Berry's phase.

Let us consider the phase that the wave function of a particle acquires as it travels around the ring structure of radius $r$. As described in the opening paragraph, in the presence of an external, perpendicular magnetic field $B_{ext}$, the particle picks up an AB phase $\delta_{AB}=2\pi (\Phi_{ext}/\Phi_{0})$, where $\Phi_{ext}=B_{ext} \cdot \pi r^{2}$ is the magnetic flux enclosed by the ring. This $\delta_{AB}$ phase leads to the well known AB oscillations of the resistance at a frequency of $\Phi_{ext}/\Phi_{0}=\pi r^{2}/(h/e)$. In the FT spectra of Fig. 2(a), the main peak observed at a frequency of 181 T$^{-1}$ corresponds to a ring radius of 0.488 $\mu$m, consistent with the size of our ring. For a system with SO interaction, however, an additional (Berry's) phase comes about due to the field $B_{eff}$. 
\begin{figure}
\centering
\includegraphics[scale=0.36]{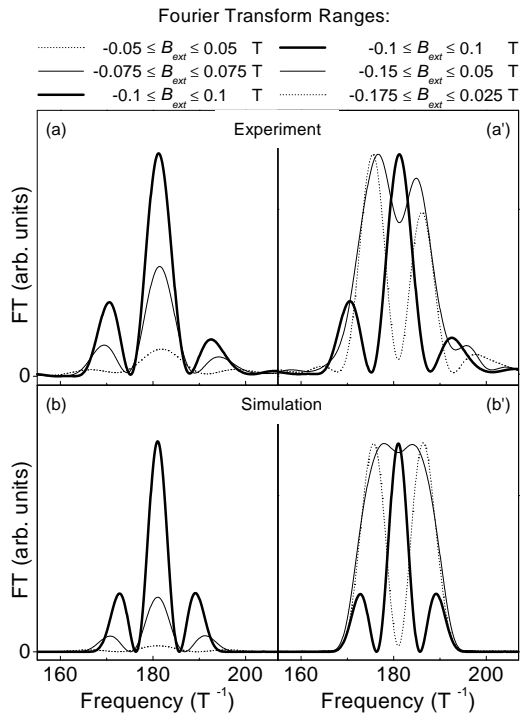}
\vspace{0.1cm}
\caption{Fourier transform (FT) spectra of the Aharonov-Bohm oscillations for different ranges of magnetic field as indicated. Top panels (a) and (a$'$) show FT spectra of experimental data taken with a gate bias of $V_{g}$ = $-0.686$ V. The bottom panels (b) and (b$'$) show the results of simulations based on Eq. (5) of the text with $B_{eff}=0.55$ T, $\Delta k$ = 2.05$\times 10^{7}$ $m^{-1}$.}
\label{Fig2}
\end{figure}
As shown schematically in Fig. 1(d), the spin of the particle travelling around the ring precesses around the net magnetic field ({\bf $\vec{B}_{ext}$+$\vec{B}_{eff}$\/}). In this situation the total phase acquired by the particle is determined by the angle ($\theta$) between the net field and the normal to the plane. Depending on the particle spin, this phase is given by \cite{Meir89,Loss92,Stern92,Aronov93,Qian94,Nitta99a,Loss00}:
\begin{equation}
\delta_{(\uparrow \uparrow or \downarrow \downarrow)}=\delta_{AB} \pm \delta_{B}
\end{equation}
\begin{equation}
\delta_{(\uparrow \downarrow or \downarrow \uparrow)}=\delta_{AB} \pm \delta_{D}
\end{equation}
with:
\begin{equation}
\delta_{B}= \pi (1- \cos \theta)
\end{equation}
\begin{equation}
\delta_{D}= \pi r \cdot \Delta k \cdot \sin \theta
\end{equation}
where $\theta=tan^{-1}[B_{eff}/B_{ext}]$. The arrow notation is used here to mark the direction of the particle spin as it travels along the two arms of the rings, e.g., $\uparrow \uparrow$ means the particle moves with like spins in both arms and $\uparrow \downarrow$ means the particle moves with opposite spins. As a result, we expect the oscillatory part of the ring resistance to be proportional to the sum of four terms \cite{dBdD}:
\begin{eqnarray}
\Delta R & \propto & \cos (\delta_{AB}+\delta_{B})+\cos (\delta_{AB}-\delta_{B}) \nonumber\\ & + &  \cos (\delta_{AB}+\delta_{D})+\cos (\delta_{AB}-\delta_{D}).
\end{eqnarray}

We now show, via simulations, that Eq. (5) indeed describes well the experimental data. Figure 2(b) presents the FT of the simulated $\Delta R$, as expressed by Eq. (5) with $B_{eff}=0.55$ T, $\Delta k$ = 2.05$\times 10^{7}$ $m^{-1}$, and $r$ = 0.488 $\mu$m. These values for the parameters $\Delta k$ and $B_{eff}$ are reasonable and are consistent with our knowledge of the SO interaction and spin splitting in samples similar to the one used here \cite{Lu98b}. The qualitative resemblance of the simulated FT to the measured data in Fig. 2 is remarkable. In particular, the simulation faithfully reproduces the side peaks observed in the experimental data. Moreover, the evolution of these side peaks with the range of magnetic field over which the FT is obtained is similar for the simulated and measured data. When the FT is performed over a small range of magnetic field, only a broad peak, centered at $\pi r^{2}/(h/e)$, and very weak side peaks are observed. As the range is made larger, the side peaks become more visible and their positions shift slightly towards the central peak position. We emphasize that in all our Fourier analyses, we used the Hamming window \cite{DSP} which significantly suppresses side peaks generated due to the finite range of data \cite{Mitra93}. Therefore, the side peaks observed in our data and simulation are genuine and not artifacts of the FT.

The results shown in Figs. 2(a$'$) and 2(b$'$) provide further evidence that Eq. (5) indeed qualitatively describes the experimental data. For these spectra, we shifted the field range over which the FT was taken so that the range was no longer symmetric around $B_{ext} = 0$. As seen in these figures, in FT spectra of {\it both\/} experimental data and simulations, the central peak splits into two strong peaks and the side peaks become weak. These observations also argue against the side peaks possibly coming from an effective multi-path structure in our sample. If such structure were present so that the peaks observed in the FT spectra came from rings of different radii, we would expect the peaks not to qualitatively depend on the magnetic field range over which the FT is performed.

\begin{figure}
\centering
\includegraphics[scale=0.365]{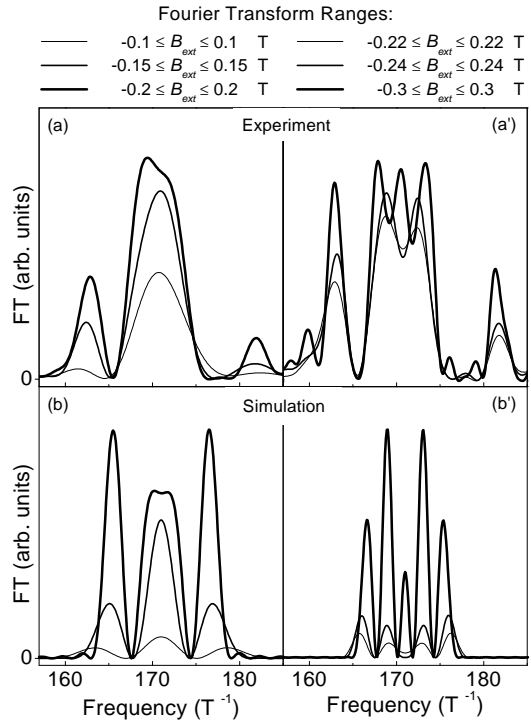}
\vspace{0.1cm}
\caption{The evolution of the FT spectra of the Aharonov-Bohm oscillations as the magnetic field range over which the FT is performed is increased. Panels (a) and (a$'$) show spectra of experimental data taken at $V_{g}$ = $-0.600$ V, while panels (b) and (b$'$) show the simulation results with $B_{eff}=0.5$ T, $\Delta k$ = 1.692$\times 10^{7}$ $m^{-1}$. Note the splitting of the main (central) peak and the reemergence of this peak in both the experimental data and simulations.}
\label{Fig3}
\end{figure}

In our measurements, we have found that, even when the magnetic field range over which the FT is performed is kept symmetric around $B_{ext} = 0$, the shapes of the FT spectra can qualitatively change if the range is made very large. To illustrate this point, in Figs. 3(a) and (a$'$), we present FT spectra of experimental data taken at a gate voltage ($V_{g}$ = $-$0.6 V), different from Fig. 2 data \cite{RingSize1}. Similar to the data of Fig. 2(a), the FT spectra exhibit a main peak and two side peaks which grow as the magnetic field range of the FT analysis increases. When the range is increased beyond $-0.2$ to $+0.2$ T, however, the shapes of the FT spectra qualitatively change.  The central peak at 171 T$^{-1}$ starts to split into two peaks and, when the range is further increased (e.g., to $-0.3$ to $+0.3$ T), the central peak reappears but is now straddled by two strong and near side peaks. Meanwhile, the original side peaks keep increasing in size so that, for the largest range shown, they are nearly as strong as the structure near the center frequency. Our simulations, shown in Figs. 3(b) and (b$'$), reveal that this evolution of the FT can indeed be qualitatively reproduced based on Eq. (5) and reasonable values of $B_{eff}$ and $\Delta k$ \cite{RingSize2}.

Some remarks regarding the values of parameters $B_{eff}$ and $\Delta k$ and how they affect the shape of the FT spectra are in order. First, to produce a spectrum with side peaks as seen in the experimental data, it is necessary to include both $\delta_{B}$ and $\delta_{D}$ phases, and to assign non-zero values to $B_{eff}$ and $\Delta k$. Second, the values of $B_{eff}$ and $\Delta k$ used in simulations of Figs. 2 and 3 were chosen to qualitatively match the experimental data. Although we do not precisely know $B_{eff}$ and $\Delta k$ for our present 2D hole system, the values we use are reasonable and realistic based on our previous measurements of the spin splitting in similar systems \cite{Lu98b}. Third, the shapes of the simulated FT spectra are in fact quite sensitive to the parameter $\Delta k$. Small changes in $\Delta k$ can change the shape of the FT qualitatively and cause the side peaks to shift in position and magnitude or even disappear. An inspection of Eq. (4) reveals that the FT spectra depend on $\Delta k$ in an approximately periodic manner: for a fixed value of $B_{eff}$, when $r \cdot \Delta k$ changes by an even integer, $\delta_{D}$ changes by 2$\pi$ and the FT of $\Delta R$ {\it vs\/}. $B_{ext}$ in Eq. (5) almost repeats. We have looked for such sensitivity and periodicity in the experimental data. Since the spin-splitting depends on the perpendicular electric field, we can in principle tune the splitting and therefore $\Delta k$ by changing the gate bias. Our data so far reveal that the shapes of the FT spectra are indeed very sensitive to the gate bias, and even hint at a periodic behavior. However, we need more systematic data to conclusively show this trend.

The overall agreement between our experimental data and simulations provides strong evidence for the observation of Berry's phase in a system with spin-orbit induced spin-splitting. Some discrepancies, however, exist. For example, the side peaks in the simulations are always equal in magnitude and are symmetric in position with respect to the main peak. In the experiments, on the other hand, we often observe some asymmetry in their positions and magnitudes. The asymmetry is puzzling \cite{IFFT}; it may be related to the finite width of the ring's arms in our sample. Future theoretical work, as well as detailed measurements of the evolution of the Fourier spectra with parameters such as gate bias, will hopefully lead to a quantitative understanding of the shapes of these spectra.

This work was supported by the NSF, ARO, DOE, and the von Humboldt Foundation. We thank B. Altshuler, A. Erbe, A. Hansen, K. Karrai, D. Loss, J.P. Lu, S. Papadakis, Y. Shkolnikov, E. Tutuc, and R. Warburton for many helpful discussions.


\begin{thebibliography}{10}
\bibitem{UCF}
Besides the AB oscillations, larger oscillations at much smaller frequencies are also observed in Fig. 1(c). The shapes of these oscillations vary with changes in the front gate voltage and they are often aperiodic and asymmetric as a function of $B_{ext}$. We do not know the origin of these oscillations; they might be related to universal consuctance fluctuations, or to geometrical resonances of electron orbits in the ring and its leads to the wider regions of the sample. We emphasize these much slower oscillations produce structures in the FT spectra at much lower frequencies, typically $\lesssim$ 20 T$^{-1}$, and do not inteerfere with the spectra of the AB oscillations that are of interest here.
\bibitem{BulkInv}
In GaAs, there is an additional inversion asymmetry resulting from its zincblende crystal structure.
\bibitem{dBdD}
The applicability of this expression to non-ballistic transport is not obvious.
\bibitem{RingSize1}
For $V_{g}$ = $-0.6$ V, we have available experimental data over a larger field range compared to Fig. 2.
\bibitem{RingSize2}
The smaller magnitude of $V_{g}$ in Figs. 3(a) and (a$'$) compared to Fig. 2 implies a smaller perpendicular electric field and therefore smaller $B_{eff}$ and $\Delta k$; also the observed main FT peak in Fig. 3(a) is 171 T$^{-1}$, implying a slightly smaller ring radius. This was taken into account for the simulations shown in Fig. 3.
\bibitem{IFFT}
The expression in Eq. (5) can be rewritten as a product of $\cos (\delta_{AB})$ and $(\cos (\delta_{B})+\cos (\delta_{D}))$, meaning that $\Delta R$ is essentially an amplitude-modulated (AM) signal with a carrier frequency equal to $\delta_{AB}$. The Fourier spectrum of such a signal consists of side-bands which are symmetrically located around the central (carrier) frequency, as seen in Figs. 2(b), 2(b$'$), 3(b), and 3(b$'$). Using standard signal processing techniques, this signal can be "demodulated" to extract the $(\cos (\delta_{B})+\cos (\delta_{D}))$ term as a direct measure of the contribution of spin-orbit interaction and Berry's phase to $\Delta R$. Unfortunately, the asymmetry of the side peaks with respect to the main peak observed in the experimental data prevents us from directly applying such a demodulation technique to our data. If we nevertheless perform an inverse Fourier transform of the structures observed at and near the central frequency, we obtain a fast oscillating signal (at the $\delta_{AB}$ frequency) whose amplitude is modulated by a slowly varying function; the latter should be closely related to the function $(\cos (\delta_{B})+\cos (\delta_{D}))$. We find a qualitative resemblance between the slowly varying function extracted from the simulations and experimental data.
\end{thebibliography}
\end{document}